\documentclass[aps,prd,showkeys,0necolumn,showpacs,preprint,11pt]{revtex4-1}
\usepackage{amsmath,amsthm,amsfonts,amssymb}
\usepackage[mathcal]{eucal}
\usepackage{mathrsfs}
\usepackage{graphicx}
\usepackage{dcolumn}
\usepackage{latexsym}
\usepackage{epsfig}
\usepackage{bm}
\usepackage[all]{xy}

\usepackage[english]{babel}
\usepackage{amsmath,amsthm,amsfonts,amssymb}
\usepackage{graphicx}
\usepackage[sort&compress]{natbib}
\usepackage{dsfont}
\usepackage{hyperref}
\usepackage{lipsum}  
\usepackage{color}
\usepackage{upgreek}
\DeclareMathOperator{\e}{e}

\DeclareMathOperator{\mx}{max}

\DeclareMathOperator{\diag}{diag}
\DeclareMathOperator{\cte}{cte}

\usepackage{hyperref}
\hypersetup{
    colorlinks,
    citecolor=blue,
    filecolor=red,
    linkcolor=blue,
    urlcolor=blue
}

\begin{document}
\title{Graviton KK resonant mode in the correction of the Newton's law from 6D braneworlds}

\author{J. C. B. Ara\'{u}jo}
\email{julio@fisica.ufc.br}

\author{D. F. S. Veras}
\email{franklin@fisica.ufc.br}

\author{D. M. Dantas}
\email{davi@fisica.ufc.br}

\author{C. A. S. Almeida}
\email{carlos@fisica.ufc.br}

\affiliation{Departamento de F\'{i}sica, Universidade Federal do Cear\'{a} (UFC), Campus do Pici, Caixa Postal 6030, 60455-760, Fortaleza, Cear\'{a}, Brazil}
 
 \begin{abstract}
In this work, we derive an expression for the correction in the Newton's law of gravitation due to the gravitational Kaluza-Klein states in a general thick string-like braneworld scenario in six dimensions. In order to analyze corrections to Newton's law we study the gravity fluctuations in a $3$-brane placed in a transverse resolved conifold and use suitable numerical methods to attain the massive spectrum and the corresponding eigenfunctions. Such braneworld model has a resolution parameter which removes the conical singularity. The correction has an exponentially suppressed mass term and depends on the values of the eigenfunctions and warp factors computed at the core peak of the brane. The spectrum is real and monotonically increased, as desired. However, the resolution parameter must assume moderate values to have physically acceptable states. Moreover, the trapped massless mode regains the 4D gravity and it is displaced from the origin, sharing similar profile with the energy density of brane for small values of resolution parameter. Finally, for the singular conifold, we found that a non-first eigenstate is a resonant mode. Such excited state is the largest contributor to corrections in the Newtonian potential.

\end{abstract}


\maketitle 


\section{Introduction}
\label{Sec_Intro}

The braneworld hypothesis has changed the way we understand the Universe introducing the idea that it may be thought as a hypersurface embedded in a higher dimensional bulk space-time. The braneworld concept emerged from string theory \cite{Horava-Witten} and unification models \cite{ADD} and yields good explanations to several fundamental phenomena in High Energy Physics such as the hierarchy problem \cite{RS1, ADJ}, the dark matter origin \cite{ADDK}, the small value of the cosmological constant \cite{CosmologicalProblem} and the cosmic acceleration \cite{CosmicAccel}.

The seminal papers of Lisa Randall and Raman Sundrum (RS) \cite{RS1, RS2} assume an Anti-de Sitter spacetime ($AdS_5$) through a warped product between a $3$-brane and a single extra dimension. As a result, the hierarchy problem can be explained by means of an exponential decreasing factor between the Planck scale and the weak scale along the extra dimension \cite{RS1}. Furthermore, the bulk massive gravitons are responsible for a small correction to the Newtonian potential of order $\mathcal{O}(r^{-2})$ \cite{RS2}. Thenceforth, several works were carried out enhancing the RS models, such as proving the stability of the geometrical solution \cite{GoldbergerWise-Radius}, providing a physical source for the geometry \cite{Csaki-UniversalAspects, Gremm, Bazeia-BlochBrane, German, 5Dthick}, allowing the Standard Model fields to propagate in the bulk \cite{GoldbergerWise-BulkFields, Huber-Shafi, Kehagias}, and extending the model to higher dimensions \cite{Cohen-Kaplan, Gregory, Vilenkin, GS}.

The extension of the RS models to six dimensions with an axial symmetry was proposed by Gherghetta and Shaposhnikov (GS) in the Ref. \cite{GS}, wherein the transverse manifold is a circle. In the $AdS_6$ GS \textit{string-like scenario}, the problem of mass hierarchy is solved without any requirement of fine tuning between the bulk cosmological constant and the brane tension \cite{GS}. Furthermore, the Kaluza-Klein (KK) excitations generate corrections to the Newtonian potential of order $\mathcal{O}(r^{-3})$. Moreover, the string-like models has the advantage of trapping free gauge fields \cite{Oda1} and minimally coupled Dirac fermions \cite{Liu1}. Besides, in the Lorentz invariance violation context, the string-like defect with a bulk dependent cosmological constant can yield a massless four-dimensional graviton \cite{LIV-stringlike}, which is not possible in the thin 5D model \cite{Rizzo}.  

Although the thin string-like model has prominent advantages over the domain-wall models, it does not satisfy all the regularity conditions inside the core of the defect and the dominant energy condition as well \cite{Tinyakov-Zuleta}. However, these issues are overcome in the thick string-like models \cite{Giovannini, Torrealba-Gravity, T2, Conifold-Scalar, Charuto, Julio}. The transverse manifold possesses its own internal symmetries, which exert influences on the geometrical and physical properties of the braneworld model \cite{Garriga-Football, Gogberashvili-AppleShaped}. Moreover, the $6D$ warped models have other interesting features. For instance, Refs. \cite{Gogberashvili-AppleShaped, Aguilar} suggest defects with two angular extra dimensions, wherein the angular momentum in the transverse space is correlated to the three generations of fundamental $4D$ fermions. Additionally, Ref. \cite{Frere} uses a single fermion family in $6D$ to explain the mass hierarchy of neutrinos. 

We work in this paper with an interesting six dimensional braneworld model with axial symmetry that uses a section of a resolved conifold (RC) as transverse space.  We confine the gravity in this scenario, and therefore this work complements the previous ones, namely, confinement of the scalar \cite{Conifold-Scalar} and matter fields \cite{Conifold-Gauge}.  The resulting scenario has a parameter $a$ which controls the singularity on the tip of the cone (see Ref. \cite{Conifold-Gauge} and references therein). Moreover, the geometric flow provided by the resolution parameter changes the properties of the analogue quantum potential for the KK massive states.  An important issue in the thick string-like braneworld scenarios concerns with  the core of the brane, which has its maximum displaced from the origin \cite{Giovannini, Charuto, Conifold-Scalar, Julio}. Such quality is due to the influence of the geometric flow in the physics of the brane \cite{Charuto}.  

In this article, we are interested in study the gravity localization on a $3$-brane placed in a transverse warped resolved conifold. We derive a general expression for the correction in the Newtonian gravitation potential due to the graviton KK states in a thick string-like braneworld scenario. Furthermore, we use suitable numerical techniques to attain the graviton spectrum and the corresponding eigenfunctions. With the eigensolutions in hands, we were able to analyse the influence of the parameters of the model in the correction of the gravitational potential.




\section{Bulk geometry and physical properties}
\label{Sec_Conifold}

In this section, we will present a brief review of the most important geometric and physical properties of the resolved conifold (RC) braneworld model. The action for a $6D$ spacetime can be denoted as \cite{Oda1} 
\begin{equation}
S_{6} =\int_{\mathcal{M}_{6}}{\left(\frac{1}{2\kappa_{6}}R-\Lambda +\mathcal{L}_{m}\right)\sqrt{-g} \hspace{0.1cm} d^{6}x} \ .
\end{equation} 
Here, $\kappa_{6}=8\pi G_{N}=8\pi/{M_{6}}^{4}$, where $G_{N}$ is the gravitational constant and $M_{6}$ is the six-dimensional bulk Planck mass. Moreover, $\mathcal{L}_{m}$ is the matter Lagrangian for the source of the geometry and the bulk cosmological constant $\Lambda$ has dimension $[\Lambda]=L^{-6}=M^{6}$. From this action, the $6D$ Einstein equations are obtained as follows:
\begin{equation}
R_{AB}-\frac{R}{2}g_{AB} = -\kappa_{6}(\Lambda g_{AB} + T_{AB}) \ .
\end{equation}  
Now, consider a static and axisymmetric warped metric between the $3$-brane $\mathcal{M}_{4}$ and the transverse space given by \cite{GS}
\begin{equation}
\label{stringlikemetric}
{ds_{6}}^2 = \sigma(\rho)\eta_{\mu\nu}(x^{\zeta})dx^{\mu}dx^{\nu} + d\rho^{2} + \gamma(\rho)d\theta^{2} \ ,
\end{equation}
where $\sigma(\rho)$ is the so-called warp function, $x^{\zeta}$ are on-brane coordinates and $(\rho,\theta)$ are coordinates of the transverse manifold. The metric has axial symmetry, so that, $r \in [0,\infty$) and $\theta \in [0,2\pi]$. The function $\gamma(\rho)$ is an angular warp factor with dimension $L^{2}$. Besides, we adopt the sign convention for the Minkowski metric as $\eta_{\mu\nu} = \diag(-,+,+,+)$. 
 
Furthermore, an axisymmetric ansatz for the energy-momentum tensor may be considered as  $T^{M}_{N}=t_M \delta^{M}_N$ \cite{GS, Oda1},
which together the metric ansatz \eqref{stringlikemetric}, leads the energy density to the form \cite{GS, Oda1, Charuto}

\begin{subequations}
\begin{eqnarray}
t_0(\rho)&=&-\frac{1}{\kappa_6}\left[\frac{3}{2} \left(\frac{\sigma^{\prime}}{\sigma} \right)^{\prime} + \frac{3}{2} \left(\frac{\sigma^{\prime}}{\sigma} \right)^{2}+ \frac{1}{2} \left(\frac{\gamma^{\prime}}{\gamma} \right)^{\prime} - \frac{1}{4} \left(\frac{\gamma^{\prime}}{\gamma} \right)^{2} + \frac{3}{4} \frac{\sigma^{\prime}}{\sigma} \frac{\gamma^{\prime}}{\gamma} \right]-\Lambda \ ,
\label{tmn}
\end{eqnarray}
\begin{eqnarray}
t_r(\rho)&=&-\frac{1}{\kappa_6}\left[\frac{3}{2} \left(\frac{\sigma^{\prime}}{\sigma} \right)^{2} + \frac{\sigma^{\prime}}{\sigma} \frac{\gamma^{\prime}}{\gamma} \right]-\Lambda \ ,
\label{trr}
\end{eqnarray}
\begin{eqnarray}
t_{\theta}(\rho)&=&-\frac{1}{\kappa_6}\left[2 \frac{\sigma^{\prime \prime}}{\sigma}+ \frac{1}{2} \left(\frac{\sigma^{\prime}}{\sigma} \right)^{2} \right]-\Lambda \ ,
\label{ttt}
\end{eqnarray}
\end{subequations}
where the other $4D$ pressures are equal to energy density $t_{1}=t_2=t_3=t_0$. Hence, the energy conditions can be addressed by the analysis of equations \eqref{tmn}, \eqref{trr} and \eqref{ttt}. Another important geometric feature in these models are the so-called regularity conditions, namely \cite {Navarro, Tinyakov-Zuleta}
\begin{eqnarray}
\sigma( \rho)\lvert_{_{ \rho=0}} &  = \cte \ , \hspace{1.7cm} \sigma^{\prime}( \rho)\lvert_{_{ \rho=0}} & =0 \ , \nonumber \\
\gamma( \rho)\lvert_{_{ \rho=0}} &  = 0    \ , \hspace{1cm} \left(\sqrt{\gamma( \rho)}\right)^{\prime}\Big|_{\rho=0} & =\cte \ . \nonumber
\label{regularity}
\end{eqnarray}
These conditions were also proposed in Ref. \cite{GS}, where the first $6D$ thin string-like model was built, the so-called Gherghetta-Shaposhnikov (GS) model. However, the warp factors of the GS model do not obey their own conditions imposed, namely
\begin{equation}\label{warp-gs}
\sigma(\rho)=\e^{-c \rho} \ , \hspace{1.2cm} \gamma(\rho)={R_0}^2 \, \sigma(\rho) \ .
\end{equation}
The parameter $c^{2} = -\frac{2}{5} \kappa_{6} \Lambda$ can be obtained from the  vacuum solution of Eq. \eqref{tmn} \cite{GS}, whereas $R_0$ is an arbitrary length scale constant related to radius of compactification disc \cite{GS}. Moreover, the scalar curvature yields to a pure $AdS_6$ spacetime where $R=-\frac{15}{2} c^2$ \cite{GS}. 

On the other hand, in order to solve the regularity issues, some other models were elaborated such as the Abelian string-vortex \cite{Giovannini} and its approximate solution \cite{Torrealba-Gravity}, the Hamilton's string-cigar and the  GS smoothed version \cite{Julio}. Nevertheless, in this work, we choose the resolved conifold (RC) model because this geometry owns a very interesting additional smoothing parameter that will regulate the corrections to Newton's law. The RC braneworld uses a $2$-section of the resolved conifold as the transverse manifold \cite{Candelas, Zayas, Conifold-Scalar}, instead of the disc of GS model.  The following warp factors are considered
\begin{equation}
\sigma(\rho)=\e^{-(c \rho \, - \, \tanh \, c \rho)}, \hspace{1.2cm} \gamma(\rho)=\left(\frac{u(\rho,a)^{2} + 6 a^{2}}{6} \right)\sigma(\rho) \ ,
\label{warp-rc}
\end{equation}
where $a$ is the resolution parameter and the dimension length function $u$, which can regularize the angular factor, has the form \cite{Conifold-Gauge}
\begin{equation}
 u(\rho,a) =
\begin{cases}
 \ \rho  \  , & a = 0\\
  -i \sqrt{6} a E\Big(\arcsin \Big(\frac{i \rho}{3 a}\Big),\frac{2}{3}\Big) \, ,&  a\neq 0
\end{cases}
\label{ua}
\end{equation}
where $E$ represents the elliptic integral of the second kind.

\begin{figure}[t] 
\begin{minipage}[t]{0.45 \linewidth}
                \includegraphics[width=\linewidth]{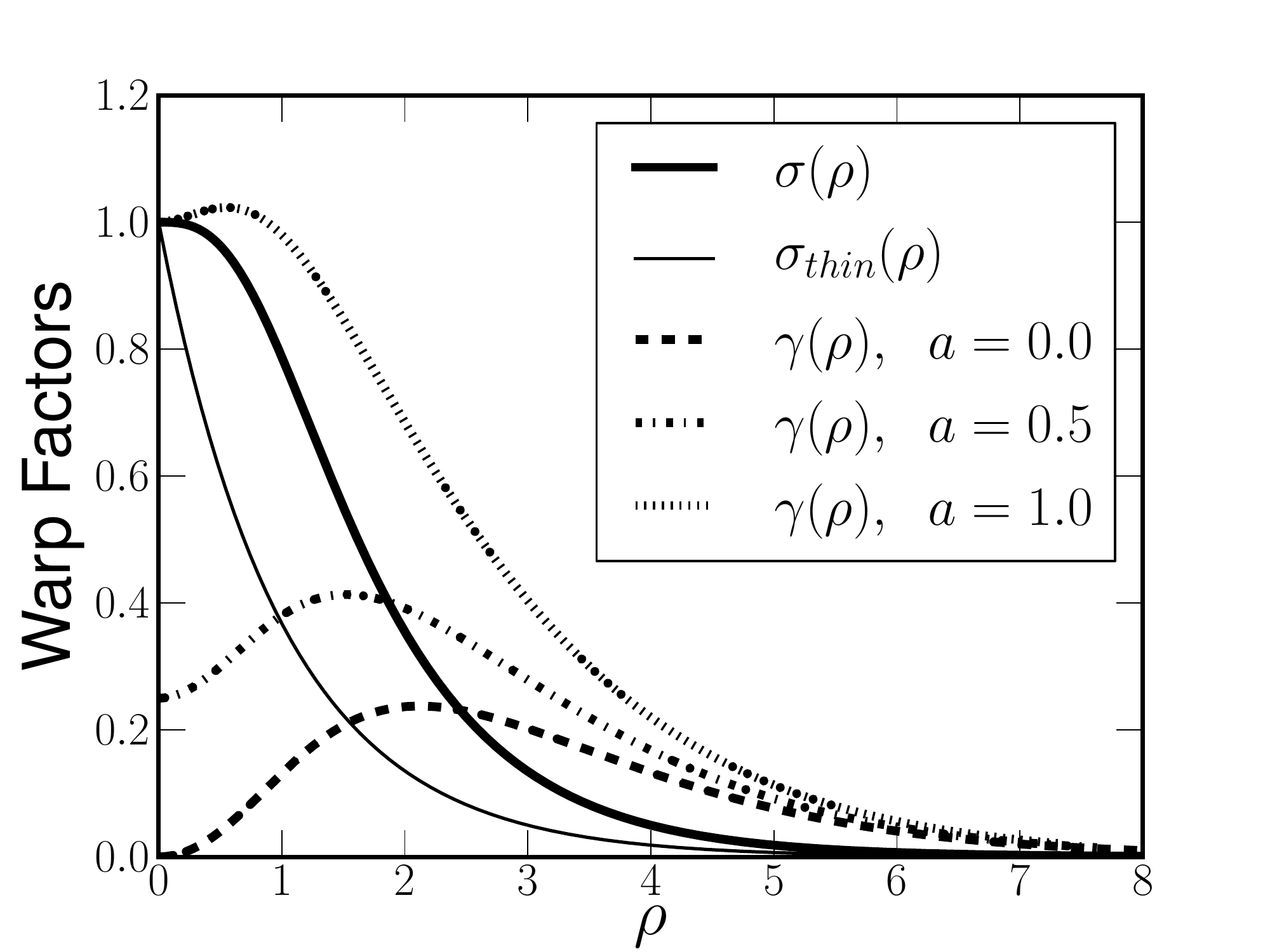}
                \caption{Warp factors $\sigma(\rho)$ and $\gamma(\rho)$ for $c = 1.0$.}
                \label{Fig_WarpFactors}
\end{minipage}
\qquad
        ~ 
\begin{minipage}[t]{0.45 \linewidth}
                \includegraphics[width=\linewidth]{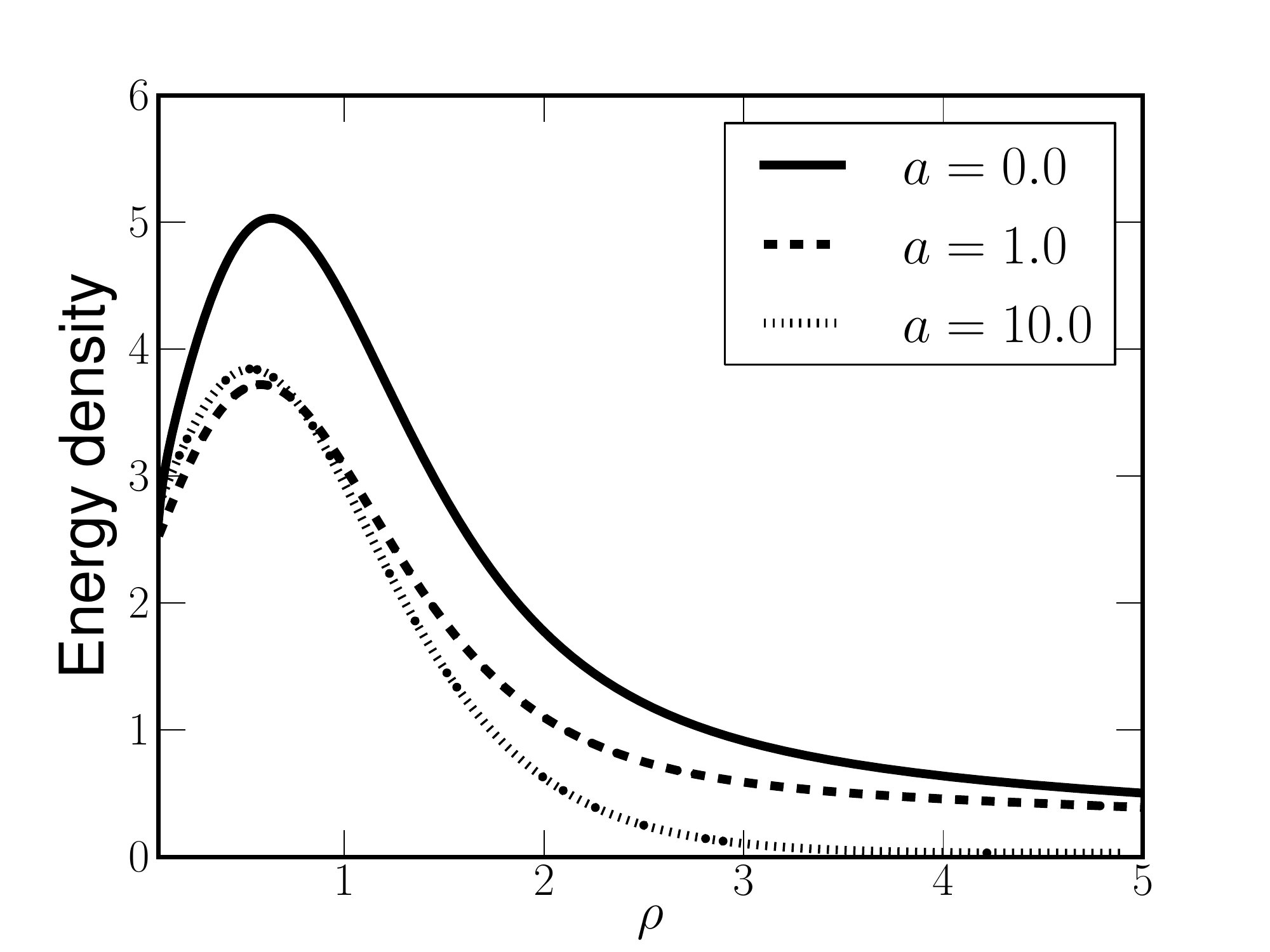}
                \caption{Energy Density $t_0(\rho)$ for different values of $a$ with $c=1$.}
                \label{Fig_Energydensity}

\end{minipage}
\end{figure}

We plot in Fig. \ref{Fig_WarpFactors} the warp factors $\sigma(\rho)$ and $\gamma(\rho)$ for $c = 1.0$ and different values of $a$. The thin string case is embedded. The warp function $\sigma(\rho)$ was first proposed in the \textit{string-cigar} braneworld \cite{Charuto}. Note that it recovers the usual thin string-like exponential behavior asymptotically. Moreover, differently from the thin string-like models, the warp function satisfies all the regularity conditions at the origin \cite{Conifold-Gauge}. Therefore, models with this warp function can be realized as a near brane correction to the thin string-like models \cite{Charuto, Julio}.

The angular ansatz given in Eq. \eqref{warp-rc} presents noteworthy features at the origin. Note that $\gamma(0) = a^2$, therefore the resolution parameter removes the conical singularity at the origin \cite{Charuto}. Furthermore, for $\rho = 0$, the $4$-brane is a $\mathcal{M}_5$ whose metric is ${ds_5}^2 = \eta_{\mu\nu}(x^{\zeta})dx^{\mu}dx^{\nu} + a^2d\theta^{2}$. Therefore, the geometric flow of the resolved conifold, driven by the resolution parameter, leads to a dimensional reduction $\mathcal{M}_6 \rightarrow \mathcal{M}_5$. The string-like dimensional reduction  $\mathcal{M}_6 \rightarrow \mathcal{M}_4$ is achieved for $a = 0$. Such features are of substantial relevance in the localization of the spin-$1$ and spin-$1/2$ fields \cite{Conifold-Gauge, Charuto-Fermions}. 


It is important to highlight that in the RC model we can regulate the shape of the energy-momentum tensor by manipulating the resolution parameter as well as the $c$ parameter (related to the bulk cosmological constant). This interesting feature is absent in most string-like models, which have only the $c$ parameter \cite{GS, Giovannini, Charuto,Julio}. Before we study the energy-momentum tensor in  Eq. \eqref{tmn} for the RC warp factors given by Eqs. \eqref{warp-rc}, we present firstly the plot of the energy density in Fig. \ref{Fig_Energydensity} for different values of $a$.

Note that the maximum of the energy density is shifted from the origin. Such feature is frequently observed in thick string-like braneworld models \cite{Giovannini, Charuto, Julio}. This displacement in the core peak of the thick branes has important influences in the massive modes of the gravitational \cite{Charuto-Gravitons}, gauge \cite{Charuto-Gauge}, fermionic \cite{Charuto-Fermions, davi2} and exotic elko \cite{Davi-ExoticElko} fields. Note from the Fig. \ref{Fig_Energydensity} that the position of  the maximum of the energy density has not great variations with the resolution parameter. Despite of this particular feature, the resolution parameter is of great importance in the mass spectrum, which has a direct influence on corrections to the gravitational potential as we will present in details in the Sec. \ref{Sec_NumericalResults}. 

\begin{figure}[t] 
\begin{minipage}[t]{0.45 \linewidth}
                \includegraphics[width=\linewidth]{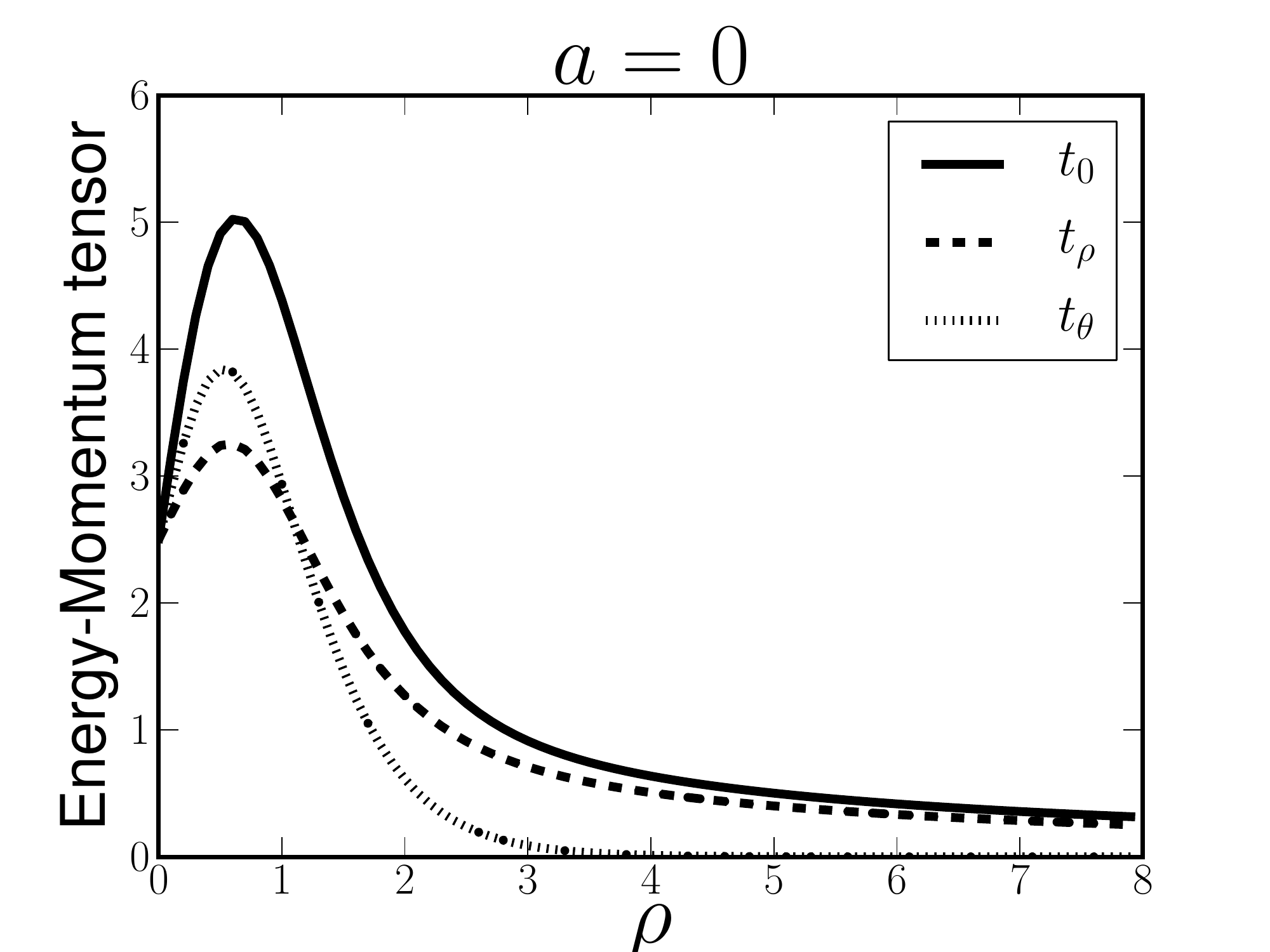}
                \caption{Energy-momentum tensor components for non-resolved conifold $a=0$ ($c = 1.0$).}
                \label{Fig_stress-0}
\end{minipage}
\qquad
        ~ 
\begin{minipage}[t]{0.45 \linewidth}
                \includegraphics[width=\linewidth]{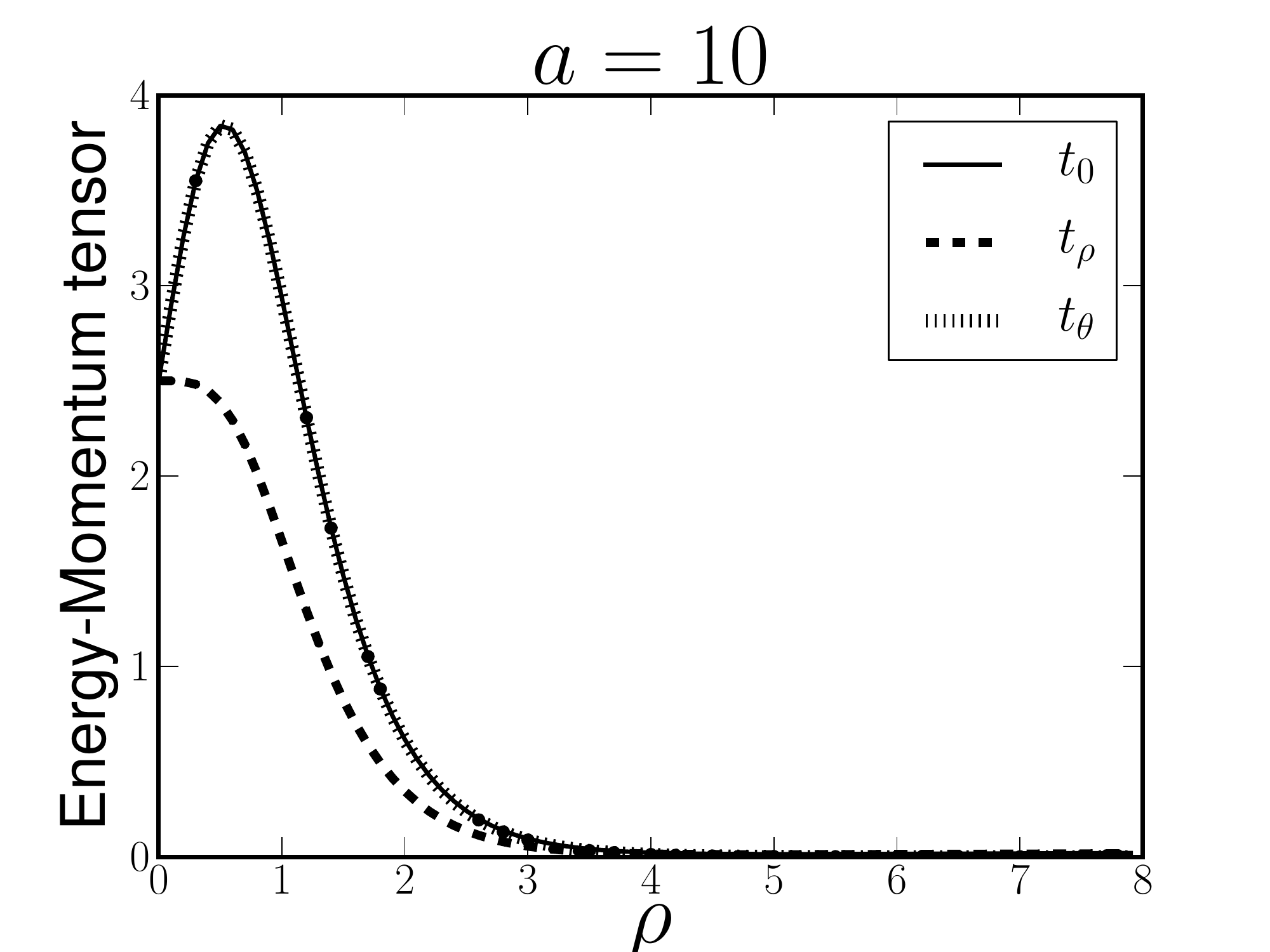}
                \caption{Energy-momentum tensor components for resolved conifold $a=10$ ($c = 1.0$).}
                \label{Fig_stress-10}

\end{minipage}
\end{figure}

Moreover, in order to analyze the energy conditions, we plot all components of equations \eqref{tmn}, \eqref{trr} and \eqref{ttt} in Fig. \ref{Fig_stress-0} and in Fig. \ref{Fig_stress-10}, for $a=0$ and $a=10$, respectively. From these figures it is observed that all the components of the tensor are always positive and hence the null and the weak energy conditions hold. Likewise, both the strong energy condition ($t_0+t_r+t_{\theta} \geq 0$)  and the dominant energy condition ($t_0\geq t_r$ \ \text{and} \ $t_0\geq t_{\theta}$) are always verified for all cases of the parameter $a$. From Fig. \ref{Fig_stress-0} we conclude that the dominant energy conditions holds for $a=0$, as also verified in Ref.  \cite{Giovannini}. Additionally, as can be seen from Fig. \ref{Fig_stress-10}, for the resolved case the growth of resolution parameter $a$ decreases continually the shape of density energy. However that decreasing never is inferior to the value of the angular pressure, reaching the equality only when $a\to\infty$. This equality is verified in models such as the analytical case of topological Abelian string-vortex of Ref. \cite{T2}, where the angular regular conditions are not satisfied as well. Although we have no explicit form for the Lagrangian in the
RC action, these peak displacements in the energy-momentum tensor indicate that may exist a phase transition for the vortex scalar fields regulated by the resolution parameter. We will back to the discussion of this RC property in $5D$ scenarios in the Sec. \ref{Sec_NumericalResults}.


\section{Metric perturbations} 
\label{Sec_Localization}

We will now study the gravity fluctuations on a $3-$brane placed in a transverse resolved conifold (RC). It was shown recently that the scalar, gauge and fermionic fields are trapped in such braneworld model \cite{Conifold-Scalar, Conifold-Gauge}. Besides, the study of gravity localization is quite relevant since it provides phenomenology implications \cite{RS2, Csaki-UniversalAspects}. This analysis will be accomplished in the Sec. \ref{Sec_Correcao} through corrections in the Newton's law.

Consider a small perturbation $h_{\mu\nu}$ in the background metric in the form
\begin{equation}
{ds_{6}}^{2} = \sigma(\rho; c)(\eta_{\mu\nu}+h_{\mu\nu})dx^{\mu}dx^{\nu}+d\rho^{2}+\gamma(\rho; c,a) d\theta^{2} \ .
\end{equation}
Imposing the transverse-traceless gauge $\nabla^{\mu} h_{\mu\nu}$ = $0$, the linearization of the Einstein equations yields to the following equation for the gravitational perturbation \cite{GS, Giovannini, Charuto}:
\begin{equation}\label{gravity-eqm}
\partial_{A} (\sqrt{-g_{6}} \, g^{AB} \partial_{B} h_{\mu\nu} ) = 0 \ .
\end{equation}
Performing the well-known Kaluza-Klein decomposition
\begin{equation}
h_{\mu\nu} (x^{\zeta}, \rho , \theta) = \sum_{n=0}^{\infty} \tilde{h}^{(n)}_{\mu\nu}(x^{\zeta})\phi_{n}(\rho) \e^{i l \theta},
\end{equation}
where $n$ and $l$ are integers, the index $n$ labels the mass values and the index $l$ the wave-number. Imposing the mass condition $\square_{4} \tilde{h}_{\mu\nu} (x^{\zeta}) = {m}^{2} \tilde{h}_{\mu\nu} (x^{\zeta})$, the radial modes satisfy the equation
\begin{equation}
\phi^{\prime\prime}(\rho) + \left( 2\frac{\sigma^{\prime}}{\sigma} + \frac{1}{2}\frac{\gamma^{\prime}}{\gamma} \right)\phi^{\prime}(\rho) + M^{2}(\rho)\sigma(\rho)\phi(\rho) = 0 \ ,
\label{sl-gravity}
\end{equation}
where the primes denote derivatives with respect to $\rho$. The factor $M^{2}(\rho) = \left(m^{2} -  \frac{\sigma(\rho)}{\gamma(\rho)}l^2\right)$ is an effective mass containing orbital angular momentum contributions $l$. 
 
Due to the axial symmetry and in order to guarantee the self-adjointness of the differential operator in Eq. \eqref{gravity-eqm}, we adopt the following boundary conditions on $\phi_{m}$ \cite{GS, Oda1, Charuto, Julio}:	
\begin{equation}\label{cond1}
\phi'_{n}(0) = \phi'_{n} (\infty) = 0 \ .
\end{equation}
Furthermore, these modes satisfy the following orthogonality condition:
\begin{equation}\label{cond2}
\int_{0}^{\infty} \sigma(\rho) \sqrt{\gamma(\rho)} \left[ \ \phi^{*}_{n_1}(\rho) \phi_{n_2}(\rho) \ \right]d\rho = \delta_{n_1 n_2} \ .
\end{equation}

For $M^{2}(\rho) = 0$ (gravitational massless mode), a solution for Eq. \eqref{sl-gravity} satisfying the boundary conditions \eqref{cond1} is a constant $\phi=\phi_0$ \cite{GS}. Hence, from the orthonormality condition \eqref{cond2}, we have:
\begin{equation}
\phi_0 =  \left(\int_0^{\infty} \sigma(\rho)\sqrt{\gamma(\rho)} \ d\rho\right)^{-\frac{1}{2}}.
\label{ModoZero}
\end{equation}
Since that for both GS and RC models the warp factors are exponentially decreasing, the constant solutions are finite and hence the zero mode is localized. For the GS model, the solution is $\phi_0 = \sqrt{\frac{3c}{2R_0}}$. 

For $M^2(\rho) \neq 0$ (massive modes), the differential equation \eqref{sl-gravity} with the warp factors given by Eq. \eqref{warp-rc}, becomes
\begin{equation}
\phi^{\prime\prime}(\rho) - \left( \frac{5}{2}c\tanh^{2} (c\rho) - \frac{1}{2} \frac{u (u ^{2} + 9 a^{2})^{\frac{1}{2}}}{(u^{2} + 6 a^{2})^{\frac{3}{2}}} \right)\phi^{\prime}(\rho) + \left(m^{2} - \frac{\sigma(\rho)}{\gamma(\rho)}l^2\right) \e^{(c \rho - \tanh c \rho)} \phi(\rho) = 0\  .
\label{sl-rc}
\end{equation}
In the limit $\rho \rightarrow \infty$, we have	
\begin{equation}
\phi^{\prime \prime}(\rho) - \frac{5}{2}c \  \phi^{\prime}(\rho) + m^{2} \e^{(c \rho)} \phi(\rho) = 0\  ,
\end{equation}
which presents the same form of the GS model \cite{GS} for a re-scaled mass $m^2 \rightarrow m^2/\e$. Thus, the thin string-like behaviour is recovered asymptotically, as expected from thick string-like braneworlds \cite{Charuto, Julio}. 

Note further that the angular eigenvalue $l$ induces a degeneracy in the mass spectrum. Therefore, hereinafter we will deal with $s$-wave solutions. The $s$-wave solution in the thin limit has the analytical solution \cite{GS}
\begin{equation}
\phi_{m} (\rho) = \e^{(5/4) c \rho} \left[  C_{1} J_{5/2} \left(\frac{2 m}{c} \e^\frac{c \rho}{2} \right) + C_{2} Y_{5/2} \left(\frac{2 m}{c} \e^\frac{c \rho}{2} \right)\right] \ ,
\label{mass-r-gs}
\end{equation}
which diverges.  Hence, there are not massive bound states in the resolved conifold braneworld model, as well.


\section{Corrections in the Newtonian potential}
\label{Sec_Correcao}

Another approach in the study of the gravitational massive modes consists in turn the Kaluza-Klein equation \eqref{sl-gravity} into a Schr\"{o}dinger-like one. Such procedure provides the study of the phenomenological implications of the  braneworld hypothesis via corrections in the Newton's law of gravitation \cite{RS2, Csaki-UniversalAspects}. Here, for the first time, it is obtained an expression for a general thick braneworld scenario in order to calculate corrections in the Newton's law.

Hence, the transformation of coordinate $z(\rho) =\int_{0}^{\rho}{\sigma^{-\frac{1}{2}}(\rho')} \, d\rho'$ provides a conformally plane metric as ${ds_{6}}^2 = \sigma(z) (\eta_{\mu\nu} dx^{\mu}dx^{\nu} + dz^{2} + \beta(z) d\theta^{2})$, where $\beta(z) = \gamma(z)/\sigma(z)$. Moreover, performing  a change of the dependent variable as
\begin{equation}
\phi_n(z) = K(z)\Psi_n(z) \ , \hspace{0.75cm} K(z)=\sigma^{-1}(z)\beta^{-\frac{1}{4}}(z) \ ,
\label{change2}
\end{equation}
we turn the Eq. \eqref{sl-gravity} into the following Schr\"{o}dinger-like equation for the $\Psi_n(z)$ function
\begin{equation}\label{schrodinger}
-\ddot{\Psi}_n(z) + U(z) \Psi_n(z) = {m_n}^2 \, \Psi_n(z) \ ,
\end{equation}
where the dots represent derivatives with respect to the $z$ coordinate and the analogue quantum potential $U(z)$ has the form
\begin{equation}
U(z) = \mathcal{W}^2(z)+\dot{\mathcal{W}}(z) \ , \quad \mathcal{W}(z) = \left(\frac{\dot{\sigma}}{\sigma}+\frac{1}{4} \frac{\dot{\beta}}{\beta}\right) \ .
\label{Quantum-Potential}
\end{equation} 
Note that the Eq. \eqref{schrodinger} preserves the analogue supersymmetric quantum mechanics form:
\begin{equation}
\left[\frac{d}{dz}+\mathcal{W}(z)\right]\left[-\frac{d}{dz}+\mathcal{W}(z)\right]\Psi(z) \ = \ \mathcal{Q}\mathcal{Q}^{\dagger}\Psi(z) = m^{2} \Psi_m(z) \ .
\end{equation}
Hence, the absence of tachyonic modes is guaranteed and the stability of the spectrum is ensured.
 
With the above mentioned changes of variables, the boundary conditions become
\begin{equation}
\frac{\dot{\Psi}(0)}{\Psi(0)}= -\frac{\dot{K}(0)}{K(0)} \ , \hspace{1cm}  \frac{\dot{\Psi}(\infty)}{\Psi(\infty)}= -\frac{\dot{K}(\infty)}{K(\infty)} \ .
\label{cond1z}
\end{equation}
The orthogonal condition is modified to
\begin{equation}
\int_{0}^{\infty}\Psi^{*}_{m_i}(z) \Psi_{m_j}(z)dz = \delta_{ij} \ .
\label{cond2z}
\end{equation}

\begin{figure}[t]
 \centering
    \includegraphics[width=0.5\textwidth]{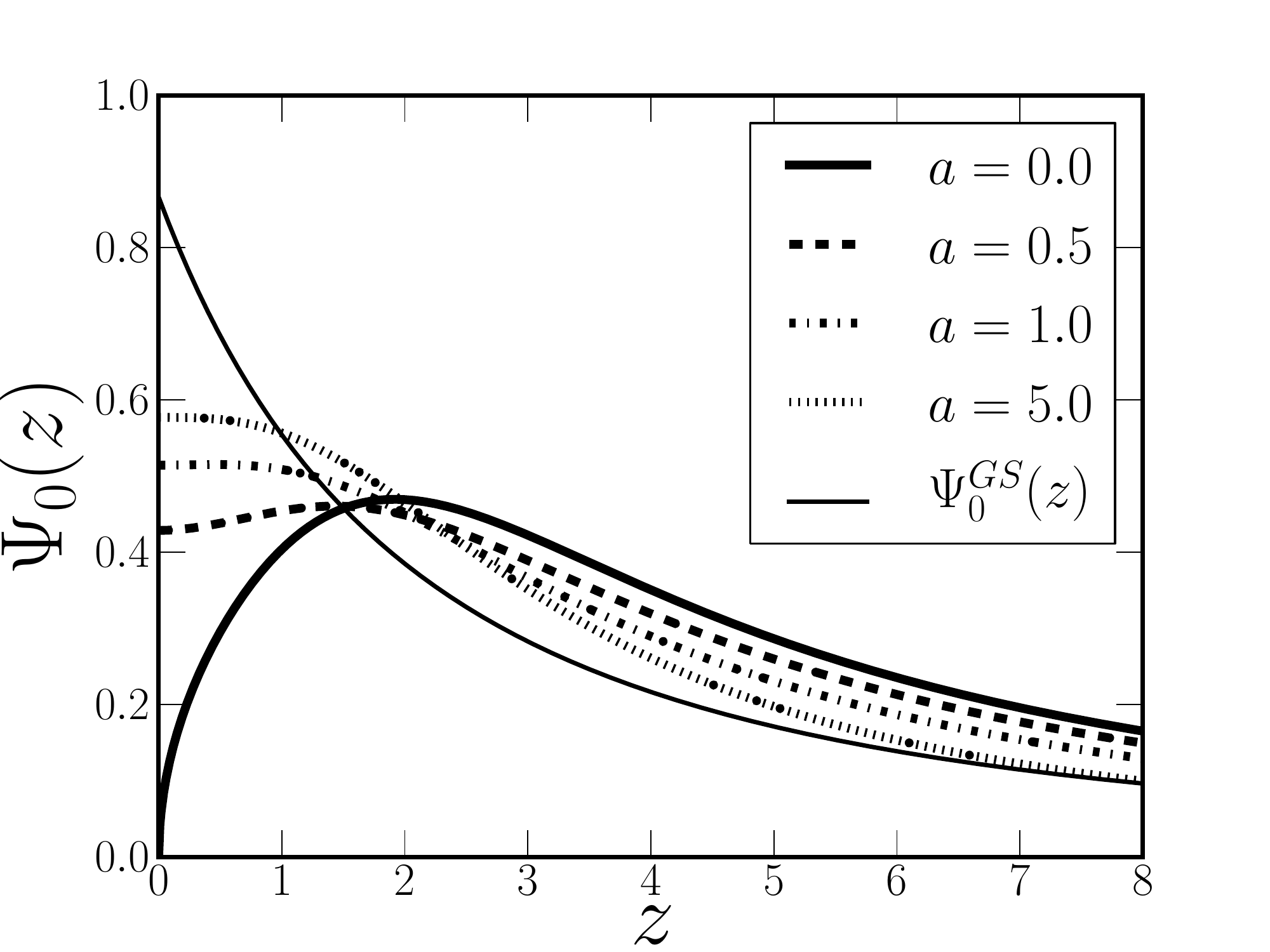}
 \caption{Gravitational massless mode in the resolved conifold braneworld. The thin string-like case is embedded. We set $c=1$.}
 \label{Fig_ModoZero}
\end{figure}

It is important to mention that the gravitational zero-mode must reproduce the four-dimensional gravity on the brane \cite{RS2, Csaki-UniversalAspects}. Hence, the solution for $m = 0$ of the Eq.  \eqref{schrodinger} is 
\begin{equation}
\Psi_0(z)=C \sigma(z)\beta^{\frac{1}{4}}(z) \ ,
\end{equation} 
where $C$ is a normalization constant. For the GS model, the zero mode in the $z$ variable is analytically evaluated as
\begin{equation}
\Psi_0(z)=\sqrt{\frac{24}{c^3}}\left(z+\frac{2}{c}\right)^{-2}.
\end{equation}
We remark here also that the gravitational zero mode is always non-singular and normalizable for all values of the resolution parameter, as showed in Fig. \ref{Fig_ModoZero}. Moreover, the massless solution shares similar profile with the energy density due to the core peak displaced from the origin for small values of $a$. Such property was also observed in the string-cigar braneworld \cite{Charuto, Charuto-Gravitons} and in the Abelian-vortex braneworld \cite{Giovannini}. For great values of the resolution parameter, the graviton zero mode has its peak at the origin. This is in accordance with the GS thin-string model \cite{GS}, in the Bounce $6D$ model \cite{davi2} and the Torrealba's string-vortex \cite{Torrealba-Gravity, T2}. This aspect of resolution parameter influences the corrections to Newton's Law.

Besides, for $m \neq 0$, the solutions of Eq. \eqref{schrodinger} in the GS scenario are given by
\begin{eqnarray}
\Psi_m(z)=\sqrt{z+\frac{2}{c}} \, \Bigg[ C_1 J_{\frac{5}{2}}\Big(mz+ 2m/c\Big)+C_2 Y_{\frac{5}{2}}\Big(mz+ 2m/c\Big)\Bigg] ,
\label{massgsz}
\end{eqnarray}
where $C_1$ and $C_2$ are constants.

We will now verify if the gravitational interactions mediated by the Kaluza-Klein modes are in agreement with the four-dimensional laws of gravity. For such purpose, we will consider a minimal coupling of matter to gravity and look for the values of the coupling constants. To compute the gravitational potential between two point-like particles with masses $\mathrm{M}_{1}$ and $\mathrm{M}_{2}$ on the brane, we rewrite the Kaluza-Klein decomposition as
\begin{equation}
h_{\mu\nu} (x^{\zeta}, z , \theta) = \frac{1}{{M_{Pl}}^2}K(z) \sum_{n=0}^{\infty} \tilde{h}_{\mu\nu}^{(n)}(x^{\zeta}) \Psi_{n}(z) \e^{i l \theta}.
\end{equation} 
Hence, we have the Lagrangian
\begin{equation}
\mathcal{L_{M}}(\Phi, g^{\prime}_{MN}) = \mathcal{L_{M}}(\Phi, g_{MN}) - \sum_{n=0}^{\infty} a_{n} (z) \tilde{h}_{\mu\nu}^{(n)} (x^{\zeta}) T^{\mu\nu} (x^{\zeta}),
\end{equation}
where the expression of the gravity-matter coupling constants is given by:
\begin{equation}
a_{n}(z) = \frac{1} {2 {M_{Pl}} ^{2}} \frac{\Psi_{n}(z) \e^{il\theta}}{\sigma(z) \, \beta^{\frac{1}{4}}(z)}.
\end{equation}

We are now able to compute the static potential generated by the exchange of the zero-mode and massive KK states. Like in the case of a Yukawa interaction, it is given by
\begin{equation}
V(r) = -\frac{1}{4 \pi} \ \sum_{n=0}^{\infty}\Big\lvert a_{n}(z=\bar{z}) \Big\lvert^{2} \ \frac{\e^{-m_{n} r}}{r},
\end{equation}
where $\bar{z}$ is the position of the maximum of the energy density on the $z$ coordinate. Therefore, on the $3$-brane, the gravitational potential between two point masses, $\mathrm{M}_1$ and $\mathrm{M}_2$, will receive a Yukawa-like contribution from the discrete nonzero modes  as:
\begin{equation}
V(r) \sim -G_{N} \frac{\mathrm{M}_1 \mathrm{M}_2}{r}\left(1 + \sum_{n>0}^{\infty} \,  \frac{\Psi_{n}^{2}(\bar{z}) \e^{-m_{n} r}}{\sigma^{2} (\bar{z}) \,\beta^{\frac{1}{2}}(\bar{z})} \right)=-G_N\frac{\mathrm{M}_1 \mathrm{M}_2}{r}\left[1+\Delta(r)\right], 
\label{Eq_Correcao}
\end{equation}
Therefore, we have an expression for general thick braneworld scenario. In order to analyze the effects of the gravitational Kaluza-Klein modes in the resolved conifold scenario to the Newton's law of gravity, we need to compute the mass spectrum and the corresponding eigenfunctions using suitable numerical methods. This will be accomplished in the next section.



\section{Numerical Analysis and experimental bounds}
\label{Sec_NumericalResults}

In the previous sections we presented the main features of a resolved conifold braneworld model and the gravity localization in this scenario. Furthermore, a general expression for the correction in the gravitational potential due to Kaluza-Klein spectrum provided from a six-dimensional thick braneworld scenario was derived. We are now able to compute this correction due to the resolved conifold braneworld. Note that the leading quantities in Eq. \eqref{Eq_Correcao} are the masses $m_n$ and the wave functions $\Psi_n(z)$ which are eigensolutions of the eigenvalue problems \eqref{sl-rc} and \eqref{schrodinger}. However, due to the involved form of the Eq. \eqref{sl-rc}, such quantities can not be obtained analytically. Then, we have used suitable numerical method to solve these Sturm-Liouville problems. 

We have used the matrix method \cite{MatrixMethod} based in finite difference schemes with second order truncation error to solve the Sturm-Liouville problem \eqref{sl-rc}. The matrix method is a numerical technique used to approximate the first eigenvalues and eigenfunctions of Sturm-Liouville problems. Indeed, in braneworld models, only the small masses are of physical interest, favouring us the use of the finite differences methods. This technique was used in the braneworld context to attain the massive spectrum in five \cite{diego5, diego6} and six dimensions \cite{Charuto-Gravitons, Charuto-Gauge, Charuto-Fermions}.

\begin{figure}[b]
 \centering
    \includegraphics[width=0.5\textwidth]{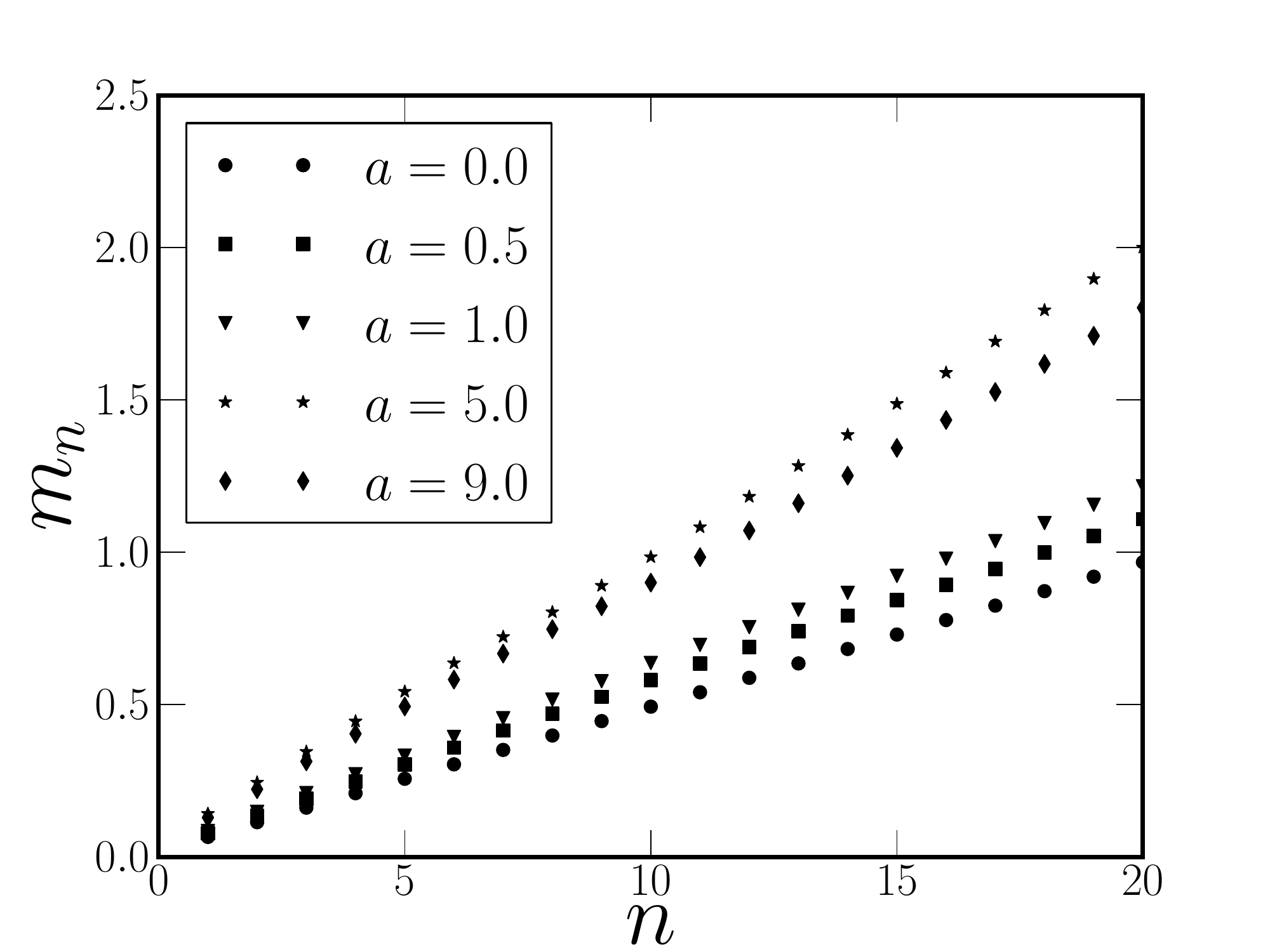}
 \caption{Gravitational Kaluza-Klein spectrum for different values of the resolution parameter $a$ and $c=1$.}
 \label{Fig_Spectrum}
\end{figure}

Here we are interested in to study the effects of the Kaluza-Klein tower in terms of the resolution parameter $a$, since the parameter related to the geometric flow $c$ acts as an energy scale \cite{Charuto-Gravitons, Charuto-Gauge, Charuto-Fermions}. Then, we solved the Eq. \eqref{sl-rc} for different values of the resolution parameter $a$ and holding $c = 1.0$ fixed. We discretized the domain $\rho \in [0.0, \, 8.0 ]$ with $N = 1\, 600$ subdivisions, in order that the constant stepsize is obtained as $h = 0.005$. 

 We plot the first twenty mass eigenvalues in Fig. \ref{Fig_Spectrum}. The spectrum is real and monotonically increasing, as expected. Note that, as the resolution parameter increases, the spectrum enhances in magnitude and the spacing between the masses enlarges. Note further that the masses exceed the limit of $m = 1.0$ for high values of $a$, and belong to a transplanckian regime (remember that $m_n \ll c = 1.0$ \cite{RS2, Csaki-UniversalAspects, Oda1, GS}). Thus, the resolution parameter must assume moderate values to have physically acceptable states.
 
With the mass eigenvalues in hands, we solve the Schr\"{o}dinger-like equation \eqref{schrodinger} with the potential function \eqref{Quantum-Potential} for the eigenvalues obtained previously. Such procedure was also followed in a sophisticated five-dimensional braneworld scenario called, hybrid brane \cite{diego5}. Since the transformation of coordinates $z(\rho)$ has no analytical expression for the warp factor in Eq. \eqref{warp-rc}, a numerical approximation was necessary to construct the analogue quantum potential. After a numerical quadrature of $z(\rho)$ for the warp factor $\sigma(\rho)$, we were able to compute the functions $\sigma(z)$ and $\beta(z)$ using cubic spline interpolation. Then, a numerical approximation for $U(z)$ was possible.

\begin{figure}[b]
 \centering
    \includegraphics[width=0.5\textwidth]{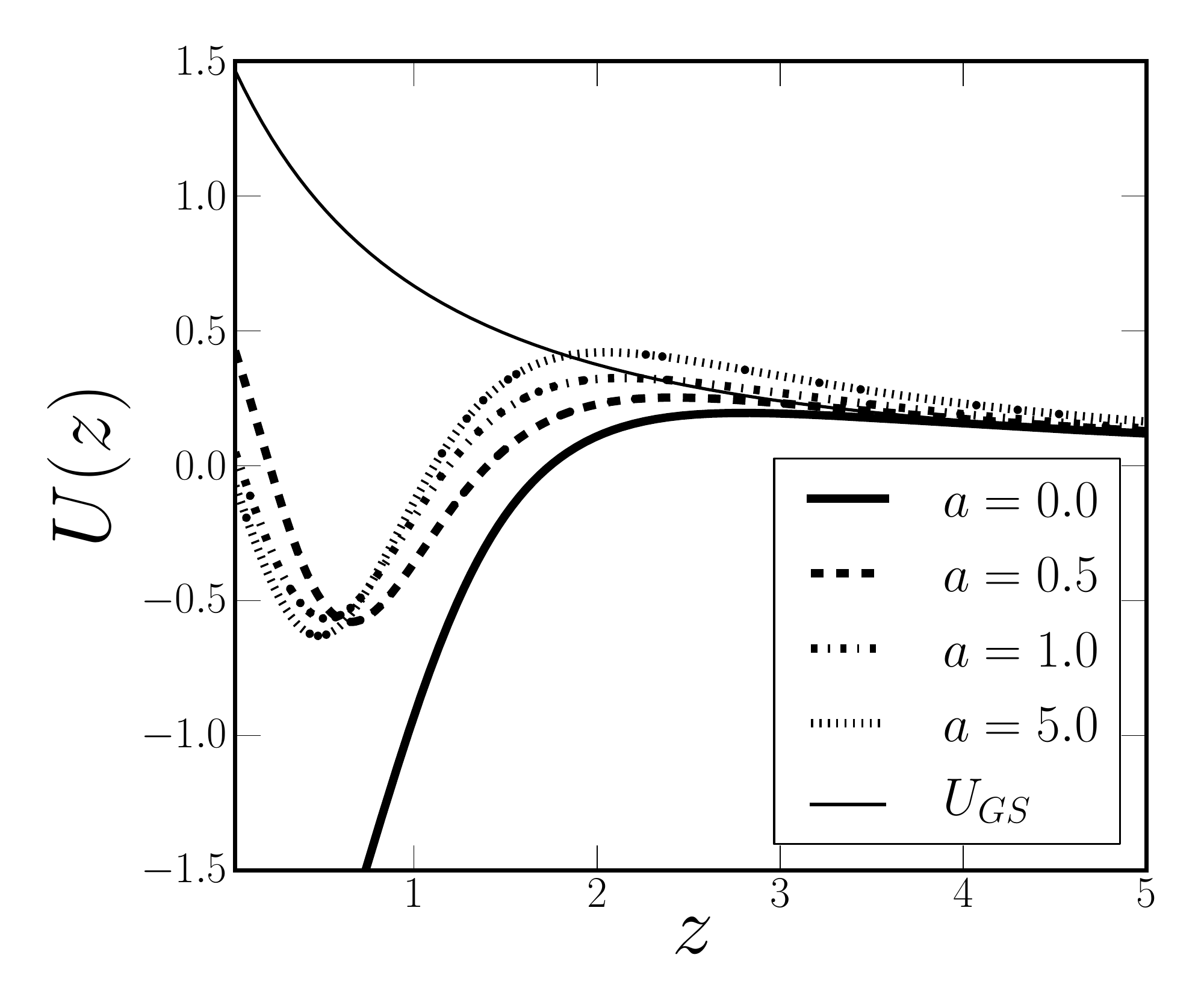}
 \caption{Analogue quantum potential for the gravitons in the Resolved Conifold braneworld scenario. The thin string-like case $U_{GS}(z)$ is embedded. We set $c=1$.}
 \label{Fig_Potential}
\end{figure}
We plot the analogue quantum potential $U(z)$ for the resolved conifold in Fig. \ref{Fig_Potential} for different values of $a$ and for $c = 1.0$. The thin string-like case is embedded. In the new coordinate $z$, the domain is stretched to $z \in [0, \, 65]$. Note that the resolution parameter removes the singularity at the origin for the RC model. Such feature was also verified in the case of scalar and matter fields \cite{Conifold-Gauge}. For $a=0$, the analogue quantum potential has a singularity at the origin. Such behaviour is also found in the string-cigar model \cite{Charuto}. Moreover, the depth of the potential well and the height of barrier are smoothed by the resolution parameter with $a\neq0$, which has no major changes for $a>5.0$.

\begin{figure}[hb] 
\begin{minipage}[t]{0.45 \linewidth}
                \includegraphics[width=\linewidth]{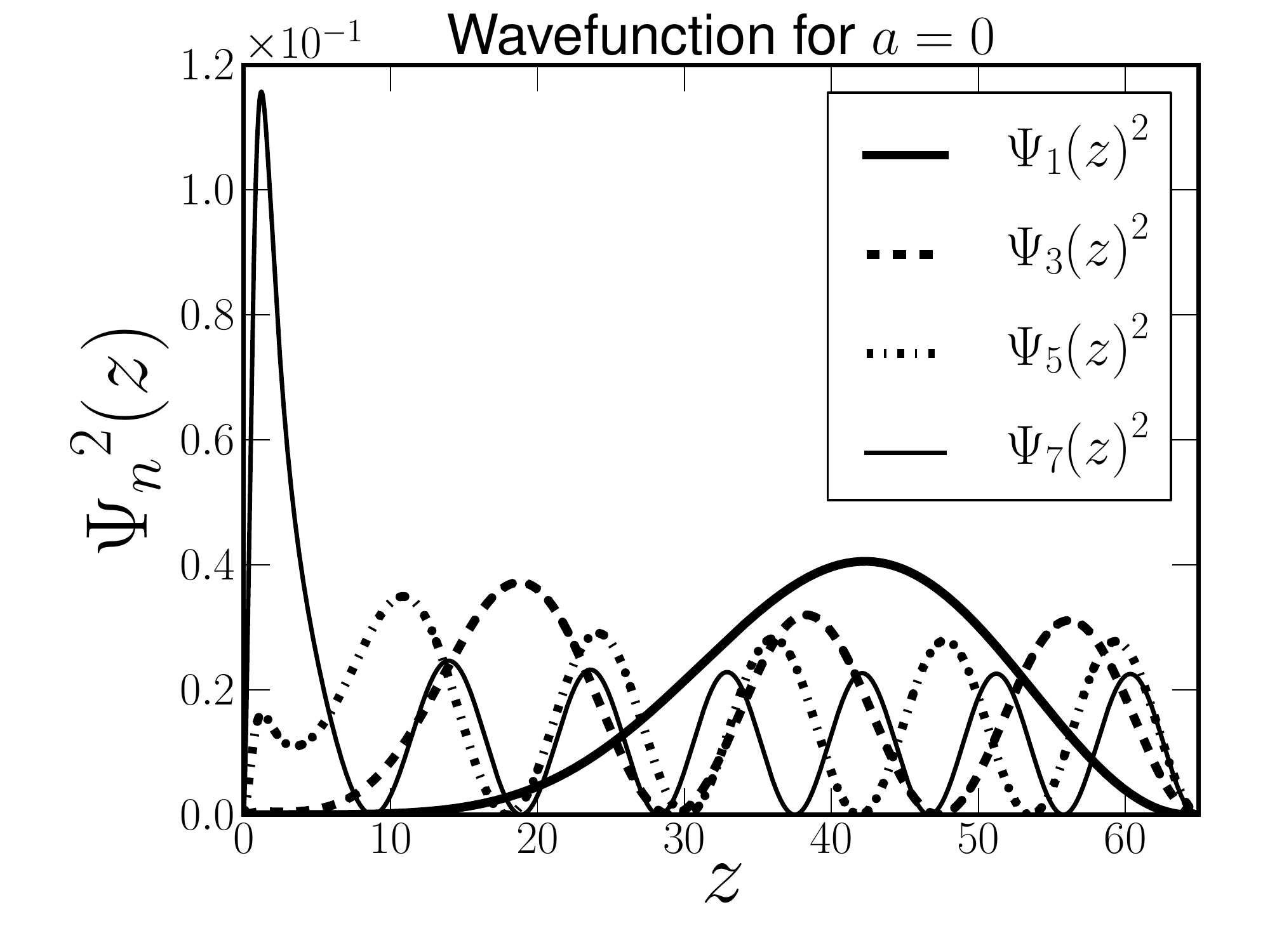}
                \caption{Normalized graviton wavefunctions for the Conifold braneworld scenario ($a = 0$) and $c=1$. The seventh eigenstate has a resonant profile.}
                \label{Fig_FuncaoDeOnda-a0}
\end{minipage}
~,\qquad
        ~ 
\begin{minipage}[t]{0.45 \linewidth}
                \includegraphics[width=\linewidth]{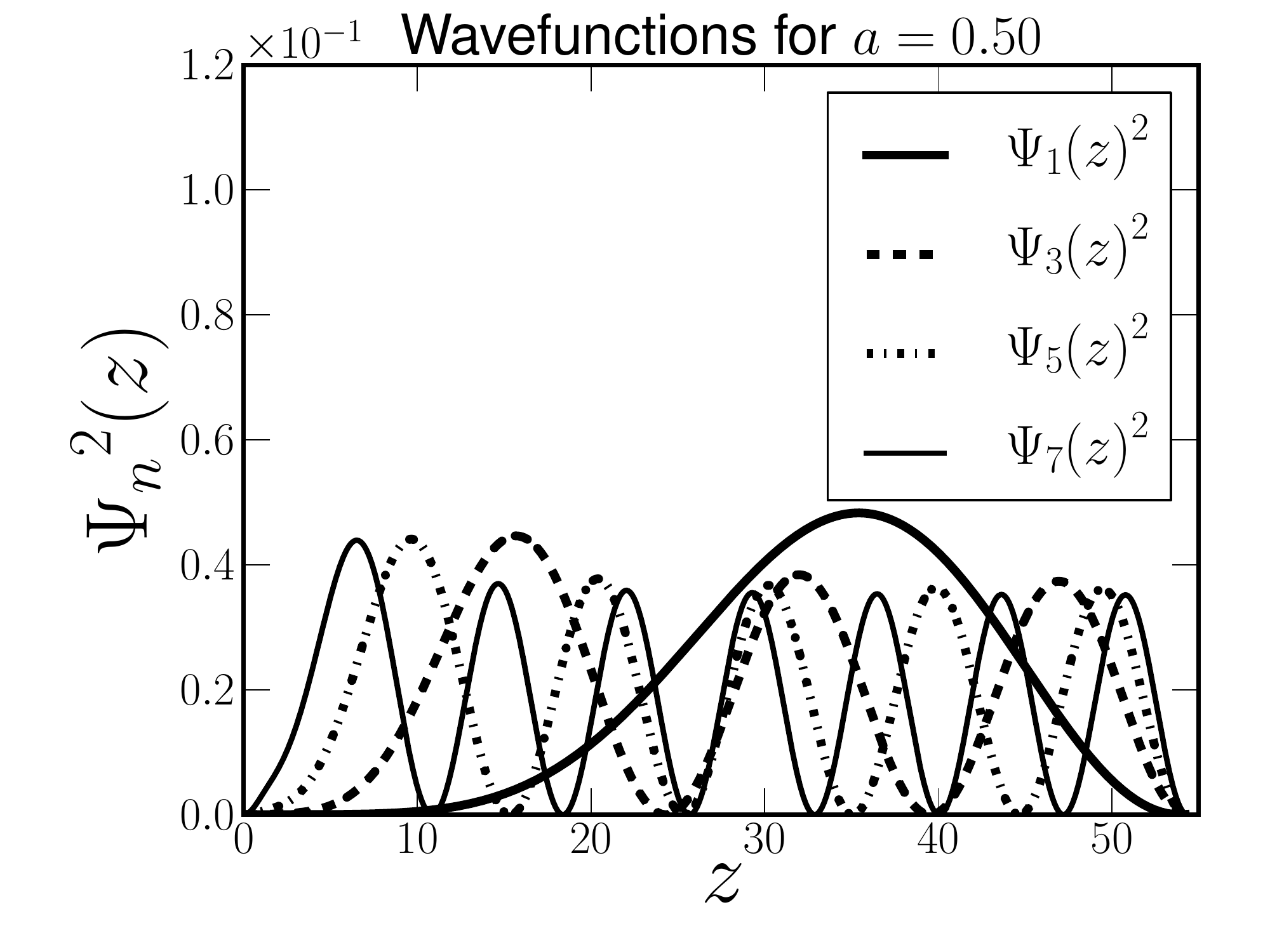}
                \caption{Normalized graviton wavefunctions for the Resolved Conifold braneworld scenario ($a\neq 0$) and $c=1$.} 
                \label{Fig_FuncaoDeOnda-a05}

\end{minipage}
\end{figure}

In order to analyze how the resolution parameter affects the gravity phenomenology of the resolved conifold braneworld, we solved the Schr\"{o}dinger-like equation using the Numerov algorithm \cite{Numerov} with the mass eigenvalues obtained previously as solution of the Sturm-Liouville problem \eqref{sl-gravity}. We plot in Figs. \ref{Fig_FuncaoDeOnda-a0} and \ref{Fig_FuncaoDeOnda-a05} the numerical solutions for $a = 0$ and $a = 0.5$, respectively, for different mass eigenvalues. A noteworthy result concerns to the seventh wavefunction for the singular conifold ($a = 0$), which has a resonant profile (massive wavefunctions having large amplitude near the brane \cite{Chineses-Ressonance}). The mass of this particular state is $m_7 = 0.351145$, which is in accordance with the limit that a resonant mass must be $m^2 < \mx (U)$ \cite{Csaki-UniversalAspects, Csaki-Quasilocalization}. In the present case, $\sqrt{\mx (U)} = 0.444009$. It is important to mention that graviton resonant states were also found in another thick string-like brane scenario with conical singularity, the string-cigar model \cite{Charuto-Gravitons}. Furthermore, the resolved case does not present resonant modes. We have solved the radial equations for several values of $a$.


This behaviour involving the existence or not of resonance for a variation in the parameter $a$, together with the displacement of the maximum in the energy-momentum tensor in Fig.  \ref{Fig_Energydensity}, reminds us about an interesting result presented in the configuration of extra dimension scalar fields in $5D$. In references \cite{W1,W2} the localization of various fields are performed in two versions of sine-Gordon models, namely the usual sine-Gordon version, which has a single kink solution, and the so-called Double sine-Gordon model, with a double kink solution. As a matter of fact, a very interesting result is presented in the double sine-Gordon scenario, which can supports a double-wall structure responsible for a splitting on the matter energy density \cite{W1,W2,W0,W3}. Moreover, the double-wall structure allows the arising of resonant KK states, which are not verified in the single sine-Gordon model \cite{W1,W2}. Similar results were also observed in the degenerate Bloch Brane \cite{W0,W3}. On the other hand, for $6D$ scenarios, an approximate solution to the topological Abelian string-vortex in six dimensions is proposed in Ref. \cite{T2}, where the vortex scalar field is a single-kink solution and the angular regularity conditions are not satisfied. Due to this, the maximum of energy is not displaced from the origin, the opposite of expected in regular models \cite{Giovannini, Charuto, Conifold-Scalar, Julio}. Further, the shape of the angular pressure is equal to the energy density, as it is denoted in the Fig. \ref{Fig_stress-10} for the RC when $a\to\infty$. In this point of view, we suspect the case where $a=0$ can represents a configuration where the scalar field for the string-vortex has a double-kink profile. Regarding the resolved case, as can be seen in the Fig. \ref{Fig_stress-10}, the displacement of energy density has a small decreasing compared to Fig. \ref{Fig_stress-0}, which it is not verified into the radial pressure. Therefore, the resolved conifold with $a\neq0$ can represent a single-kink solution to the string-vortex.

Since the main quantities that contribute to the corrections in the gravitational potential are the exponentially suppressed Kaluza-Klein masses and the corresponding eigenfunctions $-$ see Eq. \eqref{Eq_Correcao} $-$ it can be seen that the resonant state is the massive mode with largest contribution to the correction of the Newton's law of gravitation. Interestingly, in  the Ref. \cite{AssymetricRessonance1, AssymetricRessonance2}, a five dimensional asymmetric brane model exposes a resonant state performing a stronger contribution to the correction in the Newtonian potential. On the other hand, in the symmetric brane models, the first eigenstate contributes highly  for such correction \cite{Csaki-UniversalAspects, diego5}.

\begin{figure}[hb] 
\begin{minipage}[t]{0.45 \linewidth}
 \includegraphics[width=1\textwidth]{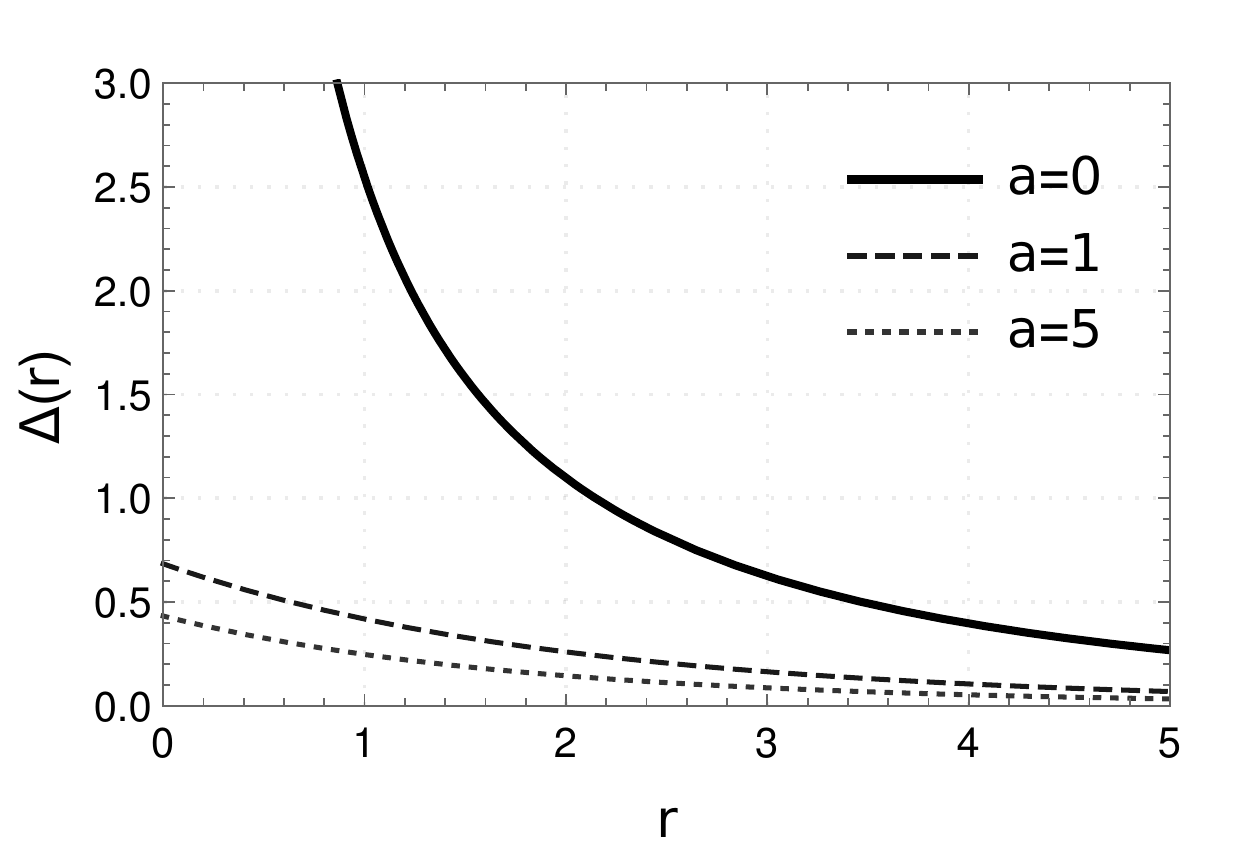}
 \caption{$\Delta(r)$: Corrections to Newton's potential in the RC model with $a=0$, $a=1.0$ and $a=5.0$ ($c=1$).}
 \label{Fig_Newton1}
  \end{minipage}
  \qquad
\begin{minipage}[t]{0.45 \linewidth}
 \includegraphics[width=1\textwidth]{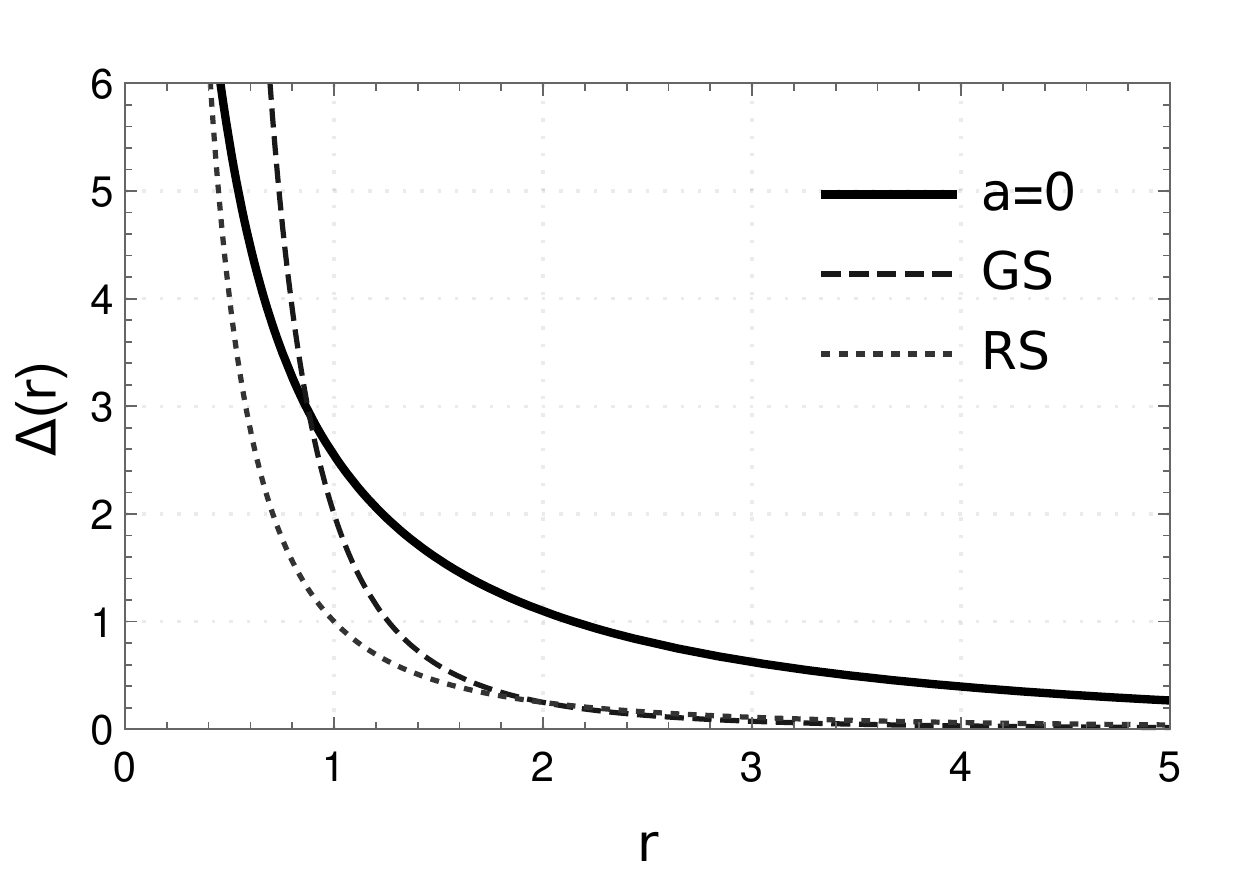}
 \caption{$\Delta(r)$: Corrections to Newton's potential in  the RC for $a=0$ compared to GS and RS models ($c=1$).}
 \label{Fig_Newton2}
  \end{minipage}
\end{figure}

From the aforementioned results, we are led to comprehend from Eq. \eqref{Eq_Correcao} that the $n = 7$ (resonant eigenstate) is the leading term of the sum for the case where $a=0$. Since for $a>0$ there is no resonant state, the corrections are very suppressed if compared to the resonant case. We express this result in Fig. \ref{Fig_Newton1}, where we note that the corrections are singular at origin when $a=0$, similar to that obtained analytically for the RS model ($\Delta(r) \approx \left(c r\right)^{-2}$ \cite{RS2,Csaki-UniversalAspects, Callin:2004py, Azam:2007ba, Eingorn:2012yu, Parvizi:2015uda, Iyer:2015ywa, Palma:2007tu, Guo:2010az})  and to the GS model ($\Delta(r) \approx \left(cr\right)^{-3}$ \cite{GS, Torrealba-Gravity, Bronnikov:2006jy}). However, the corrections has a non-singular value at origin for $a \neq 0$, which resembles the single Yukawa-Like coupling $\Delta(r)\approx \alpha \e^{-\frac{r}{\lambda}}$ (where $\alpha$ and $\lambda$ are regulate parameters) presented in references \cite{Riveros, Glicenstein, Kapner, Murata, Adelberger, Perivo}. Therefore, we also compare our numerical results for $a=0$ with the GS model and RS model showed in Fig.  \ref{Fig_Newton2}, where we note that the RC model for $a=0$ has a correction similar to RS model close to the origin, but has a slower decay when $r\to\infty$. Additionally, it is worthwhile to mention that  the graviton exchange was computed near the point where the energy of the brane is maximum, namely $\tilde{z}$, which varies little with the changes in the $a$ parameter.



\begin{figure}[hb] 
 \includegraphics[width=.5\textwidth]{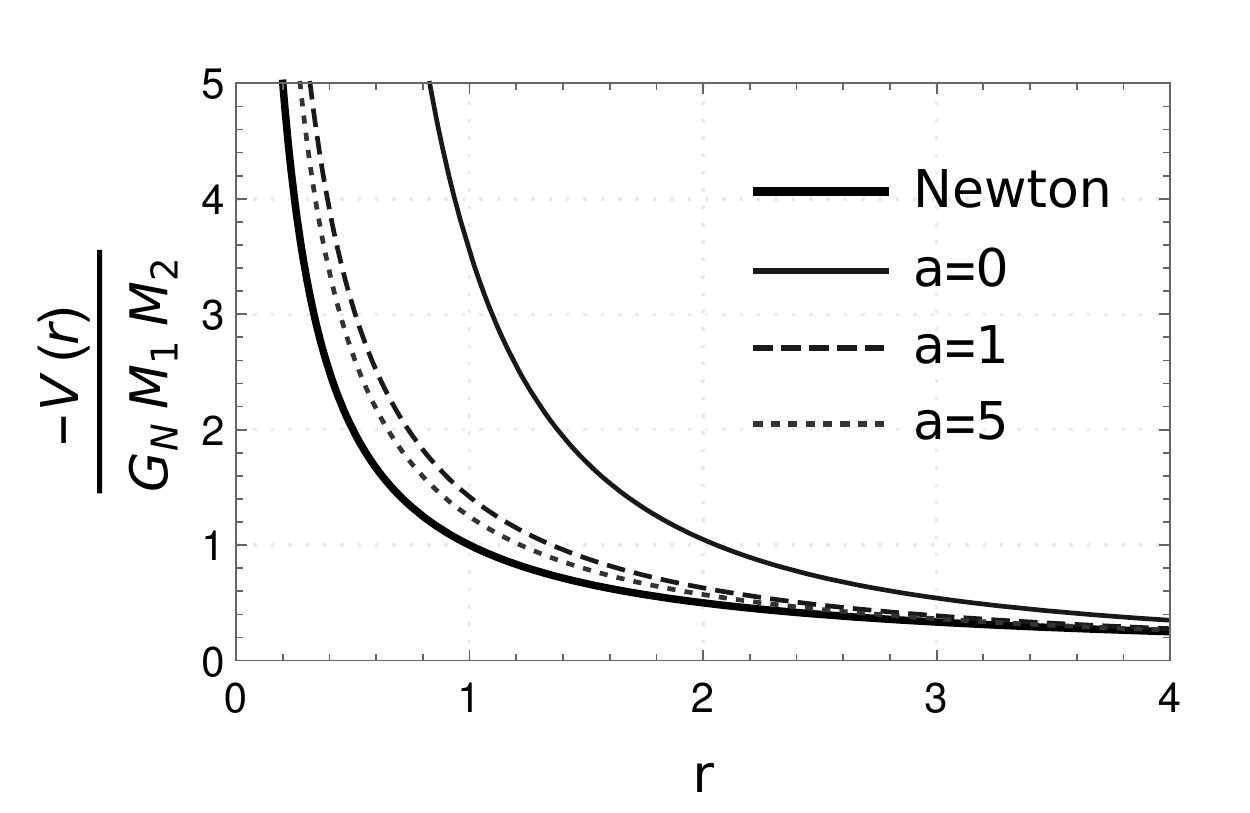}
 \caption{$\Delta(r)$: Corrected Newton's potential for RC model with $a=0$, $a=1$ and $a=5$ ($c=1$). The curve marked as Newton denotes the non-corrected potential.}
 \label{Fig_Newton3}
\end{figure}

Moreover, the corrected Newtonian potential, $V(r)$, is presented in Fig. \ref{Fig_Newton3} with different values for the parameter $a$. We note one more time that the correction performed by resonant state of $a=0$ is more expressive than the non-resonant case. In all cases, the corrections are exponentially suppressed, and the gravitational potential is slightly increased only for short distances.

 Furthermore, in order to set bounds to  RC model parameters, we can use the constraints of the solution to hierarchy problem in $6D$ \cite{GS}
\begin{equation}
M_p^2=2 \pi M^4_6 \int_{0}^{\infty}d\rho {\sigma(\rho) \sqrt{\gamma(\rho)}} \ ,  \label{hie}
\end{equation}
where $M_p\sim 10^{19}$ GeV and $M^4_6\sim 10^3$ GeV are the four dimensional  and the bulk (six dimensional) Planck scale, respectively \cite{RS1, GS}.

We must first analyze this result for the analytic GS model, applying the warp factors of Eq. \eqref{warp-gs} into Eq. \eqref{hie}. So, the relation which solves the hierarchy problem reads
\begin{equation}
\left(\frac{M_p}{M^2_6}\right)^2=\frac{4 \pi}{3} \frac{R_0}{c}  \ . \label{hie2}
\end{equation}
Alternatively, for the Randall-Sundrum model this hierarchy is explained by the relation $cL\sim 35$, where $L=\pi r_c$ is the distance between branes \cite{RS1,Iyer:2015ywa}. Therefore, the range to $c$ is very wide \cite{RS1,T2}, from a few eV \cite{Parvizi:2015uda}  until close to the Planck scale  \cite{Iyer:2015ywa}.

Finally, for the RC model we can estimate $c\sim 10^{-13}$ (or $10^6$ GeV) by numerically solving the hierarchy problem in Eq. \eqref{hie} for the non-resolved case of RC in Eq. \eqref{warp-rc} and Eq. \eqref{ua}. Hence, using this result for the GS in Eq. \eqref{hie2}, we find that $R_0\sim 10^{13}$ Planck lengths (or $10^{-22}$ m), which agrees with the bounds set for the correction to Coulomb's Law on the $6D$ string-vortex model \cite{T2}. Similarly, for the resolved case we can choose $c\sim 10^{7}$ GeV for $a\sim 10^{-22}$ m  and $c\sim 10^{8}$  GeV for $a\sim 10^{-21}$ m, among many other combinations in order to satisfy the hierarchy of equation \eqref{hie}.
%

It is worthwhile to mention that, from  another point of view, the proposition of a short range Yukawa type potential to describe the deviations from Newton's law of gravitation is widely adopted in the literature \cite{Riveros, Glicenstein, Kapner, Murata, Adelberger, Perivo}. Among many other references, we can cite initially a theoretical approach to foresee the corrections to the gravitational force by experiments based on the light and microwave solar deflection during total solar eclipses \cite{Riveros}. The gravitational microlensing observations in the force acting on photons\cite{Glicenstein} and the use of three torsion-balance experiments in a non-warped extra-dimensional scenario \cite{Kapner} are other approaches. This issue is also treated in a wide range of extensions of general relativity (GR) including $f(R)$ theories \cite{Perivo}. Furthermore,  in Randall-Sundrum model scenarios, the Ref. \cite{Callin:2004py} study high order corretion to Newton's law and some other variations in gravitational force in the $5D$ models are presented in Refs. \cite{Azam:2007ba,Eingorn:2012yu,Parvizi:2015uda}. On the other hand, supersymmetric aspects are considered in RS-like models and its implications to Newtonian potential in Ref. \cite{Palma:2007tu}.  Finally, the  Ref. \cite{Murata} brings a good updated review of several experiments of corrections to the Newton's law and its respective limits of parameters and experimental restrictions.


\section{Conclusions and perspectives}
\label{Sec_Conclusions}

In this work we have studied the gravity fluctuations in a brane placed at a transverse warped resolved conifold (RC). The RC model is a thick string-like braneworld model, whose transverse space is a $2$-section of a resolved conifold. Such brane model traps the scalar \cite{Conifold-Scalar}, gauge \cite{Conifold-Gauge}  and fermionic fields. In this work we show that the RC braneworld also traps the gravitational field. The RC model has a parameter $a$, called resolution parameter,  which removes the conical singularity. This parameter has strong influences in the geometric and physical properties of the model \cite{Conifold-Scalar, Conifold-Gauge}. We therefore, studied the effects of the resolution parameter in the phenomenological implications of the resolved conifold model via small deviations in the Newtoninan potential. We derived a general expression for the correction in the gravitational potential between two point-like sources of mass due to the gravitational Kaluza-Klein (KK) states in a thick string-like braneworld. The correction has an exponentially suppressed mass term.

The massless mode, which is responsible to reproduce the four-dimensional gravity, is trapped at the brane and is non-singular and normalizable for all the values of the resolution parameter. Moreover, the massless mode is displaced from the origin sharing similar profile with the energy density of the brane.

Using suitable numerical methods, we attained the mass spectrum of the KK graviton. The spectrum is real and monotonically increasing, as expected. However, as the resolution parameter increases, the spectrum enhances in magnitude so that for high values of $a$, the masses belong to a transplanckian regime. Thus, the resolution parameter must assume moderate values to have physically acceptable states. We also computed the analogue quantum potential. The resolution parameter removes the singularity of the potential at the origin. Nonetheless, for high values of $a$ the potential well has not significant changes.

We solved numerically the Schr\"{o}dinger-like equation for the analogue quantum potential using the mass eigenvalues for several values of the resolution parameter. For $a = 0$ (singular conifold) we found a resonant mode which is the seventh eigenstate. However, for $a\neq 0$ (resolved conifold), we have not found resonant states. Since the value of the wavefunction at the  core peak of the brane is the quantity which is used to compute the correction in the gravitational potential, the resonant state is the major contributor to this correction. A noteworthy result is that a highly excited state is the responsible for the correction in the Newtonian potential rather than the first eigenstate.

We present results for a fixed energy scale for the mass $m \ll \, c = 1.0$, such that the resolution parameter $a$ should preferably be smaller than one to have physically acceptable states. Similar results were obtained for different values of $c$. Moreover, we set experimental bounds to both parameter $c$ and $a$, based in the resolution of the hierarchy problem, which we can adjust to current experimental data. Motivated by this peculiar result, we intend for future works, to study the phenomenological consequences of the resolved conifold in the context of the Configurational Entropy \cite{Davi-Bounds,Davi-Boundsb}, computing bounds in the resolution parameter. Moreover, we will study the correction to Coulomb's Law \cite{coulomb2,coulomb2b} in thick string-like six dimensional scenarios.

\section{Acknowledgments}

The authors thank the Coordena\c{c}\~{a}o de Aperfei\c{c}oamento de Pessoal de N\'{i}vel Superior (CAPES), the Conselho Nacional de Desenvolvimento Cient\'{i}fico e Tecnol\'{o}gico (CNPQ), and Funda\c{c}\~{a}o Cearense de apoio ao Desenvolvimento Cient\'{i}fico e Tecnol\'{o}gico (FUNCAP) for financial support.

\section*{}

\end{document}